\documentclass[twocolumn, tighten]{aastex61}
\usepackage{amsmath}

\shorttitle{Interstellar communication. III. Optimal frequency to maximize data rate}
\shortauthors{Michael Hippke}

\begin{document}
\title{Interstellar communication. III. Optimal frequency to maximize data rate}

\author{Michael Hippke}
\affiliation{Sonneberg Observatory, Sternwartestr. 32, 96515 Sonneberg, Germany}
\email{hippke@ifda.eu}

\author{Duncan H. Forgan}
\affiliation{SUPA, School of Physics \& Astronomy, University of St Andrews, North Haugh, St Andrews, Scotland, KY16 9SS, UK}
\affiliation{St Andrews Centre for Exoplanet Science, University of St Andrews, UK}

\begin{abstract}
The optimal frequency for interstellar communication, using ``Earth 2017'' technology, was derived in papers I and II of this series \citep{2017arXiv170603795H,2017arXiv170605570H}. The framework included models for the loss of photons from diffraction (free space), interstellar extinction, and atmospheric transmission. A major limit of current technology is the focusing of wavelengths $\lambda<300$\,nm (UV). When this technological constraint is dropped, a physical bound is found at $\lambda\approx1$\,nm ($E\approx\,$keV) for distances out to kpc. While shorter wavelengths may produce tighter beams and thus higher data rates, the physical limit comes from surface roughness of focusing devices at the atomic level. This limit can be surpassed by beam-forming with electromagnetic fields, e.g. using a free electron laser, but such methods are not energetically competitive. Current lasers are not yet cost efficient at nm wavelengths, with a gap of two orders of magnitude, but future technological progress may converge on the physical optimum. We recommend expanding SETI efforts towards targeted (at us) monochromatic (or narrow band) X-ray emission at 0.5-2 keV energies.
\end{abstract}

\section{Introduction}
The idea to communicate with Extraterrestrial Intelligence was born in the 19th century. The first known proposal was made by the mathematician Carl Friedrich Gau{\ss} who suggested to cut a giant triangle in the Siberian forest and plant wheat inside. His motivation was that ``a correspondence with the inhabitants of the moon could only be begun by means of such mathematical contemplations and ideas, which we and they have in common.'' \citep{crowe1999extraterrestrial}. Such a passively displayed message could be reinforced following an idea by Joseph von Littrow, who suggested in the 1830s to dig trenches in the Sahara desert, again in precise geometric shapes, and fill these with kerosene; lit at night they would signal our presence \citep{van1992frontiers}.

Humans also looked for similar features on other planets. The late 19th century was a time when many great canals were built on earth, such as the Suez Canal (1869) and the Panama Canal (start in 1880). Naturally, it was thought that similar projects would be undertaken by other advanced civilizations as well, and \citet{1878MmSSI...7B..21S} announced the discovery of ``canali'' on Mars, confirmed by \citet{1904LowOB...1...59L}.

It was not long after the invention of the radio by Marconi (1897) when Nikola \citet{tesla} searched for radio signals, and detected coherent signals which he believed to originate from intelligent beings on Mars. In the optical, William \citet{pickering} suggested to send a Morse code to Mars using reflected sunlight with gigantic mirrors.

Today, we are amused about these na{\"i}ve attempts. Yet, it is enlightening to reflect on the thoughts which Phil Morrison had when writing his seminal paper on radio SETI \citep{1959Natur.184..844C}: ``And Cocconi came to me one day -- I've often reported this -- and said, 'You know what, Phil? If there are people out there, won't they communicate with gamma rays that'll cross the whole galaxy?' And I said, 'Gee, I know nothing about that (...) Why not use radio? It's much cheaper. You get many more photons per watt and that must be what counts.' And we began working on that and pretty soon we knew enough about radio astronomy to publish a paper called, 'Searching for Interstellar Communications.'\,{''} \citep{Gingerich2003}.

Morrison's statement of radio being superior is correct for the case where no tight directivity is needed, such as an omni-directional beacon. For constant power, constant detector efficiency per photon, and negligible extinction, the number of photons at the receiver is maximized for lower-energy photons. Thus, microwaves work well for an interstellar lighthouse. However, in their paper they discuss the directed communication between two nearby (10\,LY) civilizations, and the authors are well aware of the weak focusing for radio waves, because they hope for larger (200\,m class) dishes in the future. They also seek to maximize the data rate, but struggle in their bandwidth analysis: The correct framework for the photon-limited communication capacity was only recently completed by \citet{holevo1973bounds} and \citet{2004PhRvA..70c2315G,2014NaPho...8..796G} as discussed in detail in the first paper in this series \citep{2017arXiv170603795H}. Morrison's reason to disfavor optical and $\gamma$-ray communication was that they ``demand very complicated techniques'', a correct assessment in the 1950s.

Incidentally, only months later, optical communication became feasible with the discovery of the laser by \citet{1960Natur.187..493M}. It was quickly realized that the tighter beam of optical communication can increase the data rate compared to radio \citep{1961Natur.190..205S}; the authors noted the rapid development: ``No such device was known a year ago''.

To sum up, the evolution of interplanetary/interstellar communication technology migrated from planting trees, burning kerosene, reflecting sunlight, to sending electromagnetic waves in the radio, microwave and optical. Every technological step was driven by the availability of new technology, consistently considered to be the best option possible. But what if each one is an observational bias, comparable to the streetlight effect\footnote{A policeman sees a drunk man searching for something under a streetlight, and asks what he has lost. He says he lost his keys and they both look under the streetlight together. After a few minutes the policeman asks if he is sure he lost them here, and the drunk replies, ``no'', and that he lost them in the park. The policeman asks why he is searching here, and the drunk replies, ``this is where the light is''.}: people are searching for something and look only where it is easiest? 

Because of the fast technological changes, we should not focus on the ``sweet spot'' of our time, but instead derive the physical optimum. More advanced civilizations can be assumed to have developed the technology for such optimal communication.

\section{Optimal frequency}
\label{photons}
In this paper, we restrict ourselves to photons as information carriers and neglect other particles such as neutrinos or protons, as well as exotic methods such as gravitational waves, occulters or inscribed matter. These other options will be treated in another paper of the series.

We will now derive the optimum photon communication frequency within the laws of physics. The quest is as follows: Assume a civilization has two distant (e.g. $d=1.3$\,pc) locations that wish to communicate with each other. Both know of each others existence and position, they operate circular apertures as transmitter $D_{\rm t}$ and receiver $D_{\rm r}$, and they strive to maximize the number of bits exchanged (per unit time, DR) for some power $P_{\rm t}$ which might be large, but finite. Interstellar extinction between the two stations ($S_{\rm E}$) has been measured per frequency, and both stations are in space with no atmospheric losses, $S_{\rm A}=1$. Using the framework derived in \citet{2017arXiv170603795H}, we can calculate the data rate (in bits per second) as

\begin{align}
\label{eq2}
\text{DR} = \frac{S_{\rm E} P_{\rm t} D_{\rm t}^2 D_{\rm r}^2}{4 h f Q^2 \lambda^2 d^2} \times \nonumber \\  g(\eta M + (1-\eta) N_{\rm M}) - g((1-\eta)N_{\rm M})
\end{align}

where the first term has the number of photons received (per second if $P$ is in Watt), $h$ is Planck's constant ($\approx6.626\times10^{-34}$\,J\,s) and $Q\approx1.22$ is the diffraction limit for a circular transmitting telescope \citep{rayleigh} of frequency $f=1/\lambda$. 

The second term is the Holevo capacity (in bits per photon), where $\eta$ is the receiver efficiency and $g(x)=(1+x) \log_2 (1+x)-x \log_2 x$ so that $g(x)$ is a function of $\eta \times M$. $M$ is the number of photons per mode, and $N_{\rm M}$ is the average number of noise photons per mode \citep{2014NaPho...8..796G}.

\begin{figure*}
\includegraphics[width=.5\linewidth]{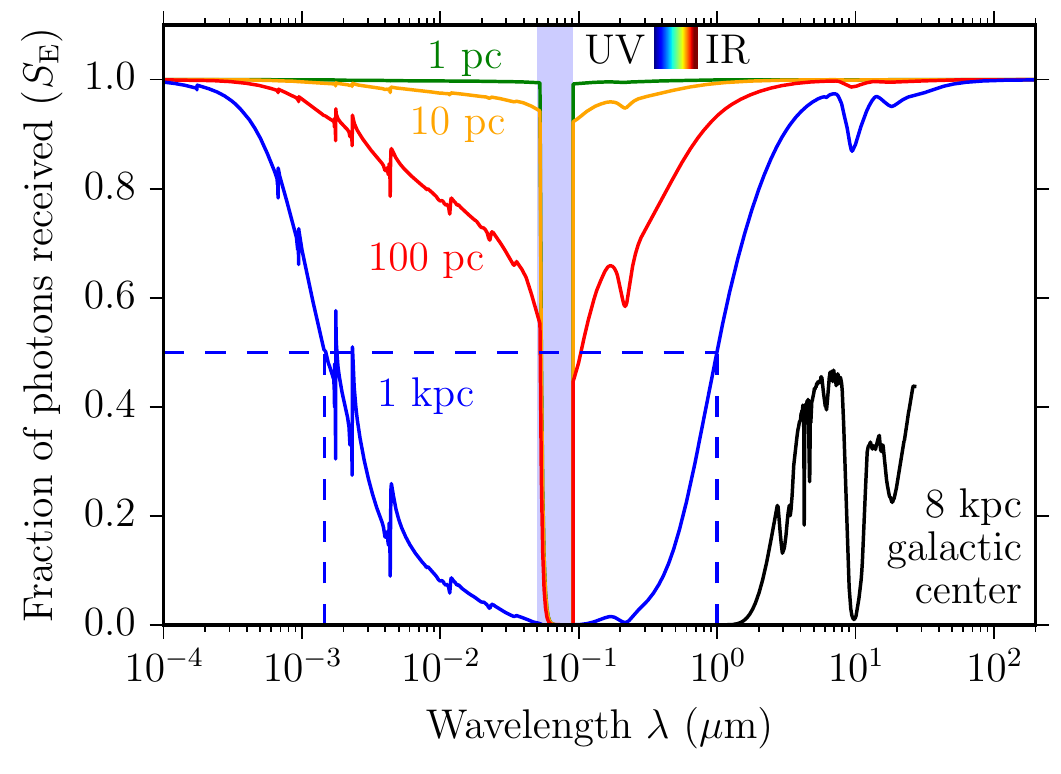}
\includegraphics[width=.5\linewidth]{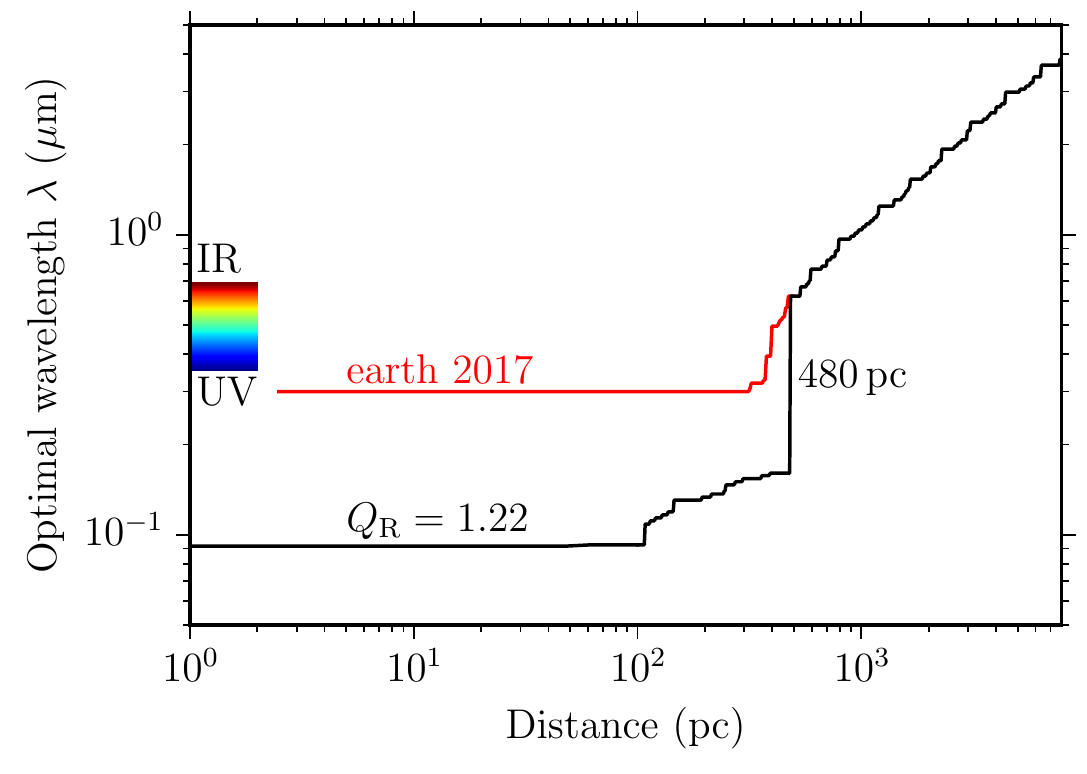}
\caption{\label{galcenter}Left: Fraction of photons that defies interstellar extinction ($S_{\rm E}$), as a function of wavelength $\lambda$, shown for different distances. The dashed blue line shows 50\% extinction for the 1\,kpc distance which occurs for $\lambda=1\,\mu$m and $\lambda=0.5$\,nm on both sides of the Lyman continuum ($\lambda\approx50\dots91.2$\,nm) which is opaque even for the closest stars due to the ionization of neutral hydrogen \citep{1959PASP...71..324A,2000ApJ...542..914W,2007eua..book.....B}. Right: Optimal wavelength for communication as a function of distance, shown for the case of low-energy photon communication ($\lambda>91.2$\,nm). Current technology (red line) is only diffraction-limited down to $\lambda\gtrsim300$\,nm, while the physical limit (black line) favors wavelengths close to the Lyman limit.}
\end{figure*}

While extinction is a function of wavelength and distance, we can neglect its influence for wavelengths far from the Lyman continuum (91.2\,nm, Figure~\ref{galcenter}) for short (pc) distances, because it is of order unity.

Equally, capacity (in bits per photon) is a function of wavelength, because $N_{\rm M}$ (the number of noise photons per mode) depends on the number of modes $M$, and on wavelength-dependent background flux. While astrophysical noise varies over several orders of magnitude by wavelength, capacity follows a logarithmic relation so that the impact is only a factor of a few as long as the flux is larger than the noise.

To build some intuition before entering a detailed discussion, we set power, distance, aperture, $S_{\rm E}$, $M$ and $N_{\rm M}$ as constant. Then, eq.~\ref{eq2} scales as

\begin{align}
\label{eq3}
\text{DR} \propto f Q
\end{align}

so that higher frequencies have a positive linear relation for higher data rates. Photon energy depends on wavelength, $E=hc/ \lambda$, which makes higher frequency photons more costly in terms of energy. Yet, the beam angle decreases linearly for higher frequencies, and thus the flux decreases quadratically for the area in the beam. The overall effect is a positive linear function so that higher frequencies allow for more photons in the receiver, all other things equal.

The factor $Q$ imposes a limit due to the mechanical quality of the machine. As we will see in section~\ref{sec:surface_quality}, there is a physical limit to $Q$ for high frequencies (small wavelengths) because mirror smoothness is limited to atomic sizes, and focusing with electromagnetic fields is inefficient (section~\ref{fel}).

\subsection{Focusing: limits from surface roughness}
\label{sec:surface_quality}
For diffraction-limited telescope mirrors or lenses, the polished surface needs to have a smoothness smaller than the wavelength \citep{rayleigh,1935lett.book.....D}. Precisely, a perfect surface would concentrate 84\% of the light into the central Airy Disk, and the remains into the diffraction rings. A ``good'' optical surface is typically assigned to a Strehl value of 80\% \citep{1894tdfa.book.....S}, resulting in 64\% of the flux inside the Airy Disk, and requiring $< \lambda/4$ peak-to-valley (pv) as well as $< \lambda/14$ root-mean-square (rms) surface accuracy \citep{2008moed.book.....S}. 

This accuracy is trivially possible at microwave frequencies, e.g. 1.4 GHz corresponds to $\lambda\approx21.4$\,cm, so that $\lambda/14\approx1.5$\,cm, surpassed by any metal dish. Parabolic optical ($\lambda=400 \dots 700$\,nm) mirrors are regularly polished to $<100$\,nm accuracy by amateurs using hand tools \citep{1984htmt.book.....T}. The physical limit for any material is set by the atomic radius of order 0.1\,nm \citep{bohr1913}. Theoretical X-Ray focusing limits stem from the fact that X-rays are scattered at the electrons of the atomic shell \citep{1948JOSA...38..766K,2004JaJAP..43.7311S}. Resulting limits have been calculated for many materials such as Be (0.028\,nm) and Os (0.034\,nm) \citep{yu1999surface}. The best suited material appears to be the one with the highest density which can be used to fabricate a reflector, and is currently believed to be osmium ($\rho=22.57$) \citep{2004JaJAP..43.7311S}.

In practice, mirrors with a surface quality at this level have been produced \citep{2001NIMPA.467.1282Y,2009JPhCS.186a2077T}. For example, the mirror of the ``Coherent X-ray Imaging instrument'' has been measured with a surface roughness of 0.57\,nm (rms) and 2.5\,nm pv \citep{2012OExpr..20.4525S}. The X-ray mirror of the XFEL instrument is specified as 0.25--0.5\,nm rms \citep{2016JOpt...18g4011B}. The mirrors of the Chandra space telescope were polished to 0.185--0.344\,nm rms, close to the size of one Ir atom (r=0.135\,nm) \citep{2012OptEn..51a1013W}.

Taking a surface roughness of 0.2\,nm rms results in a 80\% Strehl ratio for $\lambda > 2.8$\,nm. The theoretical physical limit allows for ``good'' optics for $\lambda > 0.48$\,nm using Osmium. This limit could hypothetically be extended using degenerate matter (e.g. from Neutron stars) where the electrons are compressed more tightly, but it is unclear if such a fabrication is possible, so that it is neglected here. We will instead adopt a minimum wavelength of order $\lambda=1$\,nm as the plausible physical limit, a factor of two added to the hard cut, because a reflective coating at these wavelengths will likely not be possible with the densest material (Osmium), as discussed in the following section.

\subsection{Reflectivity at nm wavelengths}
Classical parabolic mirrors at these wavelengths suffer from low reflectivity, because the refractive index of most materials converges to unity at high (keV, $\lambda\approx$\,nm) energies. This makes it difficult to focus photons efficiently and avoid absorption \citep{2008PhyU...51...57A}, and it makes lenses impossible with known materials. Current limits at $\lambda=11$\,nm (extreme ultraviolet, EUV) and normal (not gracing) incidence angle are 80\% reflectivity using exotic alloys \citep{2000ApOpt..39.2189S}. For shorter wavelengths (X-rays, below 5\,nm/250\,eV), grazing incidence mirrors (at angles of a few degrees) are common. Their disadvantage are long focal lengths and/or the need for multiple mirrors. However, technological progress in material sciences is rapid and order unity reflectivity for nm wavelengths at normal incidence angles might be feasible with yet unknown alloys.

\begin{figure}
\includegraphics[width=\linewidth]{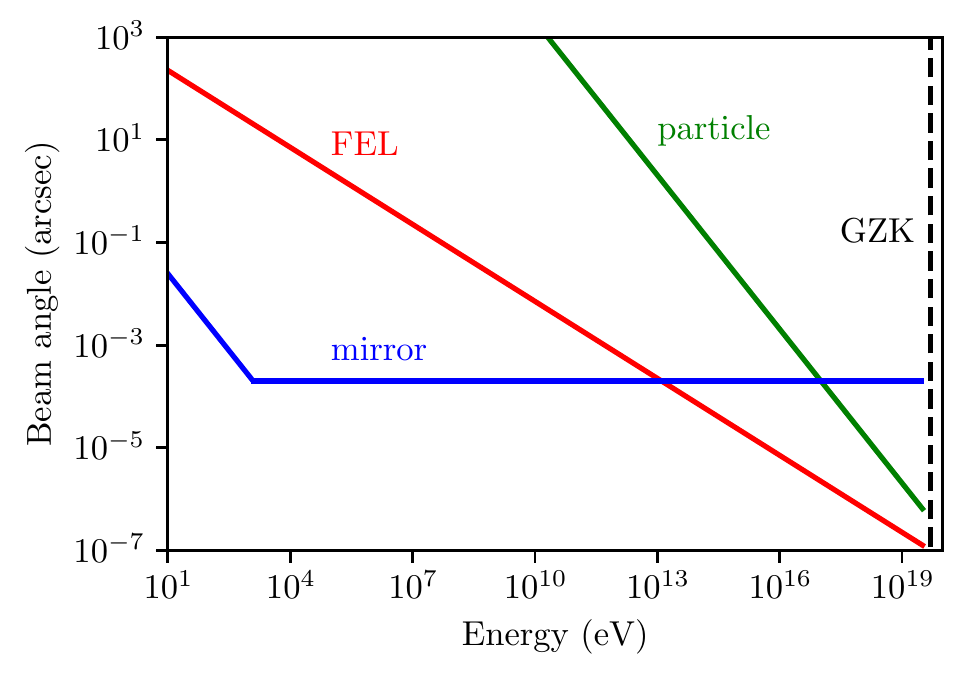}
\caption{\label{fig:beamwidth}Beam angles of meter-sized photon mirrors, FELs and particle accelerators as a function of particle energy. The mirror cutoff is due to the atomic surface roughness. The GZK cutoff (vertical dashed line) is the Greisen-Zatsepin-Kuzmin limit \citep{1966PhRvL..16..748G,1966JETPL...4...78Z} at $5\times10^{19}$\,eV$\approx8$\,J per photon.}
\end{figure}

\subsection{Focusing with electromagnetic fields instead of physical mirrors}
\label{fel}
We might then wonder whether it is possible to surpass the limitations of material science by using electromagnetic fields to focus a beam.

An elegant method to directly convert power from electron beams to electromagnetic radiation is the free-electron laser \citep{1971JAP....42.1906M}. The synchrotron radiation generated by a single electron in a beam has a characteristic angular spread of 

\begin{equation}
\theta_{\rm electron} = \frac{1}{\gamma} \approx \frac{0.1}{E\,{\rm [GeV]}}\,{\rm [rad]}
\end{equation}

where $\gamma$ is the relativistic boost factor of an electron, and $E$ is its energy \citep{2009JInst...4T5001I}. The beam angle at GeV (TeV, PeV) energies is $6^{\circ}$ (21\,arcsec, 21\,mas). For comparison, the opening angle of diffraction-limited optics is $\theta_{\rm optics} = 1.22 \lambda / D_{\rm t}$. For $D_{\rm t}=1$\,m, $\theta_{\rm electron} = \theta_{\rm optics} $ at $\lambda=82$\,nm (15 eV), a difference of $7\times10^{10}$ in energy for the same beam width. In other words, focusing particles into a beam of given width requires $7\times10^{10}$ more energy in a particle accelerator (e.g., using TeV particles) compared to a meter-sized mirror (e.g., with 15\,eV photons). This limit applies to any particle accelerator, e.g. producing Neutrinos through muon decay. It can be considered as the minimum spread since a particle beam of positive width and angular spread will only increase this angle.

Electron beams can be used to produce photons by passing through the magnetic field of an undulator. Here, the width over which constructive interference is possible is $<1/\gamma$. Such electrons with rest mass $E_{\rm 0}=511$\,keV have a relativistic Lorentz factor, e.g. at $E=6$\,GeV \citep[used at the European Synchrotron,][]{1995JPhD...28..250B}, $\gamma=E/E_{\rm 0}\approx11,742$. An electron beam passing through the magnetic field of an undulator produces photons of $\lambda \approx \lambda_{\rm 0} / 2 \gamma^2$, where $\lambda_{\rm 0}$ is the period length of the undulator \citep{2002RvMP...74..685N}. For $\lambda_{\rm 0}=28$\,mm, we get $\lambda=0.1$\,nm. The width of the photon beam is \citep{luchini1990undulators}

\begin{equation}
\theta_{\rm photon} = \sqrt{\frac{2\lambda}{L}} 
\end{equation}

where $L=N\lambda_0$ is the length of the machine. This beam is tighter than the electron beam (for the same energy of electrons and photons). It is however wider compared to a classical mirror of the same size, and it scales with the square root of the machine size and the energy, whereas the classical mirror scales proportionally (Figure~\ref{fig:beamwidth}). Consequently, the FEL beam requires TeV energies (at m-sized machines) to achieve the $\theta=0.2\,{\rm mas}/D_{\rm t}$\,(m) beam width of a meter-sized keV classical mirror. The FEL beam width can be decreased further to $0.01\,\mu$as at the Greisen-Zatsepin-Kuzmin limit \citep{1966PhRvL..16..748G,1966JETPL...4...78Z} at $5\times10^{19}$\,eV ($\approx8$\,J per photon, section~\ref{vhf}), but this is not energy efficient. Spending the same energy on a classical keV beam would deliver more photons to the receiver.

While beam forming efficiency is insufficient to compete with physical devices, wall-plug power efficiency into photons through electron beam deceleration and recycling might ultimately increase to order unity. Literature suggests possible values of 
10\% \citep{2010IJQE...46.1135S},
14\% \citep{2016PhRvS..19b0705E} or
30\% \citep{2016PhRvL.117q4801S}. 
Small (table-top) tunable x-ray lasers might soon be possible through the use of graphene plasmon-based free-electron sources \citep{2016NaPho..10...46W}.

\section{Atmospheric turbulence}
\label{atmospheric_turbulence}
The ``Cyclops Report'', often dubbed the ``SETI bible'' \citep{1971asee.nasa.....O}, claimed the superiority of microwaves over other communication methods in the 1960s. One of the reasons was atmospheric turbulence. The report argues (correctly) that ``all laser systems suffer the disadvantage of a higher energy per photon than microwave systems (...)''. It is then admitted that ``this disadvantage is partly compensated by the ease of obtaining narrow beams'', and the following calculation compares the beam widths a 22.5\,cm aperture for an IR transmitter to 100\,m (and 3,000\,m) radio telescopes. The report justifies this choice by ``atmospheric turbulence or pointing errors'', which limited beams to arcsec widths. 

This was a valid point in the 1960s -- but it is less relevant today. The technological progress has been considerable, with adaptive optics soon reaching near-diffraction limited (mas) beam widths (and $\mu$as astrometry) at $\mu$m wavelength for large (39\,m) telescopes \citep{2016SPIE.9909E..2DD}, an improvement of 2--3 orders of magnitude. Due to atmospheric absorption for $\lambda<291$\,nm, X-ray transmitters and receivers need to be placed in space.

\section{Pointing accuracy}
\label{pointing_accuracy}
The transmitter's beam needs to be pointed such that the emitted photons arrive at the receiver's position after the photons have traveled the intervening distance. This is in general non-trivial, as beam angles at the optimum wavelength $\lambda\approx$\,nm have small widths of $\theta=0.2\,{\rm mas}/D_{\rm t}$\,(m). As we will see, this beam width is typically small compared to the expected motion of receivers orbiting nearby stars.

\subsection{Pointing at planet-based receivers}

For illustration, the orbit of a planet in the habitable zone around the nearest star Alpha Centauri A or B has an apparent width of 0.17\,arcsec, about $1000\times$ larger than the beam of a meter-sized telescope.  As a result, any transmission attempting to reach distances $\lessapprox$ 1 kpc must attend carefully to pointing accuracy.

If the receiver is located on a planet sufficiently close that the beam is smaller than its orbit, the transmitting telescope needs to account for the proper motions of the star, plus the orbit of the planet \citep[as noted by][]{2016QuEle..46..966M}. Such accuracy for stellar positions and proper motions is possible with current technology. The Gaia mission will achieve astrometry at the $\mu$as level \citep{2012Ap&SS.341...31D} in the best cases. 

The transmitter must also be able to forecast the receiver's location years or even decades in advance.  Forecasts for exoplanet orbits are achievable if they are detected either using the transit or radial velocity method.  For example, if an exoplanet is detected via transit, this can be used to predict the timing/duration of its next transit to relatively high accuracy to facilitate follow-up observations, (see e.g. the Transit Predictor Service offered by the NASA Exoplanet Archive\footnote{https://exoplanetarchive.ipac.caltech.edu/include/VisibleTransitIntro.html}).  These follow-up observations are crucial, as transit observations alone are typically insufficient to confirm the planet's full ephemeris (especially the eccentricity).  This situation improves as multiple detection methods are brought to bear, and if other planets in the system are detected by e.g. transit timing variations \citep{Agol2017}.  This allows further constraints on planetary orbits by appealing to arguments based on the orbital stability of the system.  

It remains clear that any would-be transmitter attempting to contact a planet-based receiver will begin with some observational constraints on the receiver planet's orbit.  These constraints, which may initially be quite loose, will govern strategies for initial communication.  

First attempts are likely to have a deliberately larger-than-optimal beam width, to ensure receipt of the signal in the face of uncertainties regarding the receiver's position when the signal enters their star system.  It would seem sensible for any planet-based transmitter to also include details of its own ephemeris, so that a reply from the receiver can be accurately targeted.  Ephemerides sharing is likely to be a small but significant component of all interstellar communications.

\subsection{Pointing at space-based receivers}

As the optimal frequency for interstellar communications is typically absorbed by planetary atmospheres, both transmitter and receiver are likely to be space-based.  If the transmitter orbits a planet, then much of the above section applies to this case, with the added need to share the transmitter's orbital parameters as well as the planet's.  This also adds a further source of latency - the transmitter/receiver pair are unable to connect when either is occulted by their host planet.  Depending on the orbital parameters of both, transmissions will be forced to occur at specific epochs, with well-defined time intervals or windows in which to conduct the conversation.

Another option would be to place the receiver stationary with respect to the star, for example in its gravitational lens (section~\ref{lensing}).  In this scenario, the transmitter merely needs to point their beam at the star (accounting for its proper motion), rather than forecasting the orbital motion of a planet and its telescope placed on or around it.  As noted in \citet{2017arXiv170605570H}, tight beams ($\lesssim 3R_{\odot}$ at the receiver end) should not be pointed at the star directly, but instead at the Einstein ring to maximize the flux in the image plane of the lens. For reference, a $D_{\rm t}=1\,$m telescope at $\lambda=1$\,nm and $D=1.3$\,pc produces a beam width of $\approx 0.5\,R_{\odot}$. If such a beam were centered on the star, most of the flux would be lost. Therefore, pointing accuracy and proper motion calculations at sub-mas level are required for such a communication.  Again, a transmitter choosing to use the SGL would be wise to inform the receiver of this (and details regarding the lens) in its initial communiqu\'es.

\section{Extinction}
\label{photon_extinction}
As shown in \citet{2017arXiv170603795H}, interstellar extinction prohibits communication near the Lyman limit \citep{1906ApJ....23..181L,1996Ap&SS.236..285R} between $\approx 50\dots91.2$\,nm. This gap separates interstellar communication with low energy ($E<13.6$\,eV, $\lambda>91.2$\,nm) from that with high energy. The gap widens with increasing distance due to extinction.

\subsection{Short distance ($<100$\,pc)}
For short (pc) communication distances, extinction is negligible (few percent) outside of the gap. For distances up to 100\,pc, the optimum frequency in the low energy range remains close to the Lyman limit, at about 100\,nm. High energy communication is unaffected and placed at the general optimum set by physical mirrors of 1\,nm (compare Figure~\ref{galcenter}).

\subsection{Medium distance ($100\dots500$\,pc)}
The gap widens and deepens with distance, resulting in a monotonic increase of the optimum wavelength for low energy communication, from 100\,nm at 100\,pc distances to 200\,nm over 480\,pc.

Then, at $>480$\,pc distances, a considerable step follows (Figure~\ref{galcenter}) where the optimal low-energy wavelength moves from UV (160\,nm) to IR (690\,nm). The tipping point is caused by the slope of the extinction function exceeding the focusing advantages. The jitter visible in the optimal IR wavelength comes from infrared absorption bands.

\subsection{Large distance ($500\dots8000$\,pc)}
For larger distances, the optimum wavelength increases monotonically. Over kpc, extinction levels of 50\% are present at $1\,\mu$m on the low-energy side, and at 1.5\,nm on the high-energy side. For distances of 8\,kpc (galactic center), this shifts to $3\,\mu$m and 0.5\,nm, respectively. Therefore, high energy communication near 1\,nm wavelength is possible (and optimal) over all relevant galactic distances, perhaps limited in the case of extreme extinction for the innermost few pc of the galactic center.

\subsection{Very high frequencies}
\label{vhf}
For very high frequencies, the absorption of MeV and GeV $\gamma$-rays in our Galaxy is completely negligible. To absorb a MeV photon with pair production, the target must be another MeV  photon. The product of the energies of the two photons must be of order MeV$^2$. There are simply not enough MeV (or higher energy) photons traveling in the galaxy to make a significant target. Obviously, the situation is different inside a source or in particular environments, where there is a large density of X- or $\gamma$-rays. Therefore, transmittance from the galactic center to earth is near unity for $\gamma$-rays up to $10^{12}$\,eV \citep{2006ApJ...640L.155M} or $10^{13}$ eV \citep{2016PhRvD..94f3009V}. Even higher energy photons show a prominent absorption feature between $10^{13}$ and $10^{17}$\,eV mainly due to interaction with the cosmic microwave background \citep{2006A&A...449..641Z}. The survival probability over 8\,kpc is estimated as 30\% at around $10^{15}$\,eV (=1\,PeV, or $\lambda=10^{-12}$\,nm) \citep{2016PhRvD..94f3009V}. The upper bound for high energy photons is set by the Greisen-Zatsepin-Kuzmin limit \citep{1966PhRvL..16..748G} at $5\times10^{19}$\,eV ($\approx8$\,J) by slowing-interactions of cosmic ray photons with the microwave background radiation over long distances ($\approx100$\,Mpc), which has been observationally confirmed \citep{2008PhRvL.101f1101A}.

\section{Lensing}
\label{lensing}
As discussed in part II of this series \citep{2017arXiv170605570H}, the solar gravitational lens (SGL) enlarges the aperture (and thus the photon flux) of the receiving telescope by a large factor which is a function of the receiver aperture in the image plane ($d_{\rm SGL}$), wavelength $\lambda$ and heliocentric distance $z$. For a telescope with its center on the axis ($\rho=0$), the average gain is \citep[][their Eq. 143]{2017arXiv170406824T}

\begin{eqnarray}
\begin{aligned}
\label{eq_mu}
\bar{\mu}(z,d_{\rm SGL},\lambda)=\frac{4\pi^2}{1-e^{-4\pi^2 r_g/\lambda}}\frac{r_g}{\lambda} \times \\ \Big\{ J^2_0\Big(\pi\frac{d_{\rm SGL}}{\lambda}\sqrt{\frac{2r_g}{z}}\Big)+J^2_1\Big(\pi\frac{d_{\rm SGL}}{\lambda}\sqrt{\frac{2r_g}{z}}\Big)\Big\}.
\end{aligned}
\end{eqnarray}

where $r_g = 2GM_{\odot} /c^2 \approx 2,950$\,m is the Schwarzschild radius of the sun, and $J_x$ are Bessel functions. This gain is large, for example for $d_{\rm SGL}=1$\,m, $\lambda=1$\,nm, $z=600$\,au, we get $\bar{\mu}=2.91\times10^9$.

The corresponding circular aperture size $D_{\rm classical}$ of a telescope outside of the SGL compared to a telescope of size $d_{\rm SGL}$ in the SGL is
\begin{equation}
\label{aperture_size}
D_{\rm classical} = 2 \sqrt{\frac{\bar{\mu} d_{{\rm SGL}}^2}{4}} = d_{\rm SGL} \sqrt{\bar{\mu}}.
\end{equation}

The $d_{\rm SGL}=1$\,m telescope at $z=600$\,au collects as many photons as a $D_{\rm classical}=75.8$\,km (kilometer) telescope. The influence of $z$ is within a factor of $\approx2$ between $z_0=546$\,au and $z=2,200$\,au. The difference for a change in wavelength from $\lambda=1\,$nm to $\lambda=1\,\mu$m is $\approx2$\% and can therefore be neglected.

This gain is so large that despite high coronal noise the achievable data rates are higher by a factor of $10^7$ for same size telescopes, as long as the signal is not overpowered by noise. For plausible parameters, the usable regime begins at $P_{\rm t}>0.1\,$W.

\section{Data rates}
\label{data_rate}
In this section, we will calculate the limits on data rates for different distances and parameters, with power from Watt to GW and aperture sizes from $1\dots100$\,m. These parameters cover the capabilities from ``Earth 2017'' technology with moderate cost (e.g., KW power, meter-sized space telescopes) to advanced, but still plausible ``next century'' forecasts (e.g., GW power). We assume a conservative capacity of one bit per photon, which corresponds to ``Earth 2017'' technology. Due to the logarithmic influence of the number of modes on capacity, even very advanced technology will not achieve more than 10 bits per photon in practical cases. Therefore, we can use one bit per photon as an order-of-magnitude argument. As the standard value for wavelength, we use the optimal $\lambda\approx$\,nm. With these assumptions, and neglecting extinction, the data rate scales (to within 10\%) as

\begin{equation}
{\rm DR} \approx d^{-2} D_{\rm t}^2 D_{\rm r}^2 P {\rm (bits/s)}
\end{equation}

where $d$ is in pc, $D_{\rm t}$ and $D_{\rm r}$ in m and $P$ in Watt. An important lesson from this relation is that aperture and power can be traded; a pair of meter-sized telescopes at GW power delivers the same data rate as a pair of 100\,m telescopes at 10\,W power (or 10\,m telescopes at 100\,kW power).

A pair of meter-sized probes at Watt power can deliver bits per second per Watt over pc distances. This scales to 10\,Gbits/s for $D_{\rm t}=D_{\rm r}=10$\,m at MW power. Alternatively, the 1\,m Watt-scale probes can be placed in the SGL for a high (10\,Mbits/s) data rate. Given nm communication technology, it appears almost trivial to communicate significant data over short interstellar distances. 

The DR from the exemplary pair of $D_{\rm t}=D_{\rm r}=10$\,m probes at MW power decreases to 100 (1, 0.01\,Mbits/s) over distances of 10 (100, 1000\,pc). It drops to $<100$ bit/s to the galactic center given the added loss from extinction ($S_{\rm E}\approx50$\%). Upping the machinery to $D_{\rm t}=D_{\rm r}=100$\,m at GW power yields Gbits/s data to the galactic center \citep[$7.86\pm0.18$\,kpc, ][]{2016ApJ...830...17B}, and 100\,kbits/s to Andromeda \citep[$752\pm27$\,kpc,][]{2012ApJ...745..156R}.

Foreground extinction (galactic and intergalactic) towards Andromeda is surprisingly low at A(V)=0.17 \citep{2011ApJ...737..103S}, comparable to a distance through the galactic disk of 100\,pc. The beam width of the exemplary telescope is $2.5\,\mu$as at 1\,nm, and expands to a width of 1\,au at the distance of Andromeda. Such a communication is thus highly selective and targeted at one specific star (or even planet).

The data rate of this link ($D_{\rm t}=D_{\rm r}=100$\,m at GW power) drops below kbits/s (bits/s) for distances of 10 (300) Mpc. This advanced, but plausible  technology allows for communication throughout the local group of galaxies with reasonable data rates. For reference, transmitting at bits/s delivers 31\,Mbits $\rm{yr}^{-1}$, sufficient to communicate a substantial book per year.

\section{Data as a function of money}
At first approximation, wavelength can be traded linearly with power and with the squares of the aperture diameters, ${\rm DR} \propto \lambda P_{\rm t} D_{\rm t}^2 D_{\rm r}^2$. The relation is linear with the aperture surface area.

It appears plausible that any civilization, even a very advanced, will have non-zero cost for each of these variables. Therefore, ET might strive to maximize bits per time per money. When maximizing this quantity, we can examine the influence of different types of cost functions for wavelength (surface smoothness plus photon production), power and aperture size.

Building the machine (mirror) at a certain size and quality requires fixed (one-off) costs, with additional variable costs (e.g., for maintenance) per bit. The total costs over the finite lifetime of the machine can be attributed per bit. Traditional (filled-aperture) telescope cost functions are progressive for larger sizes, because the supporting structure is to be build in three dimensions to counter gravity, so that the mirror diameter follows a cubic cost function. Recent technological advances (in the microwave domain) use aperture synthesis combining many smaller dishes, so that the surface area tends more towards a linear, or slightly progressive, cost function. It is unknown if this approach can be extended to X-ray telescopes. Tiled apertures could also be used as multiple coincidence detectors, improving signal-to-noise. In all solutions, the aperture must be sized up as long as the marginal cost for an additional increase in surface area grows beyond a quadratic function. For ``Earth 2017'' technology, this would be of order meters, or at most tens of meters, in the optical; and hundreds of meters to km for radio.

For the energy production, a linear (proportional) cost function is most plausible, because power plants of a certain capacity can be added as required, with negligible infrastructure overhead. Economies of scale may even result in a regressive cost function. There will be a limit to power because of cooling requirements, which depend on the efficiency, which we neglect here.

Regarding the wavelength, the relevant metric is the surface quality of the machine and its photon production at a constant size and power. As the relation of data rate and frequency is linear, the wavelength should be decreased until the marginal cost for a further improvement increase beyond a linear relation. For example, the baseline might be a $\lambda=1\,\mu$m photon production with a laser plus the required $\approx250$\,nm (pv) surface smoothness (which is established ``Earth 2017'' technology). Then, an improvement to $\lambda=0.5\,\mu$m and $\approx125$\,nm surface smoothness (at the same size and power) must not exceed twice the cost to be beneficial. For ``Earth 2017'' technology, the optimum is around 193\,nm using an ArF (UV) excimer laser and a good mirror, or some other UV to optical laser which is commercially cheap and adequately powerful. This is two orders of magnitude away from the physical optimum.

Earth's atmosphere absorbs wavelengths $<291$\,nm so that UV and X-ray communication devices need to be located in space. Currently, this increases costs by orders of magnitude. 

In the end the question is whether the physically optimal ($\approx$\,nm) wavelength is commercially competitive, or whether e.g. a 2\,nm device can be built at less than half the cost, and the money instead invested in more power. This question can not currently be answered. However, we may argue that very advanced technology will offer ``cheap'' nm-class space-based photon emitters and mirrors. After all, the history of the laser (and other methods) show that technology fills every niche in wavelength over time, at very similar costs.

\begin{figure}
\includegraphics[width=\linewidth]{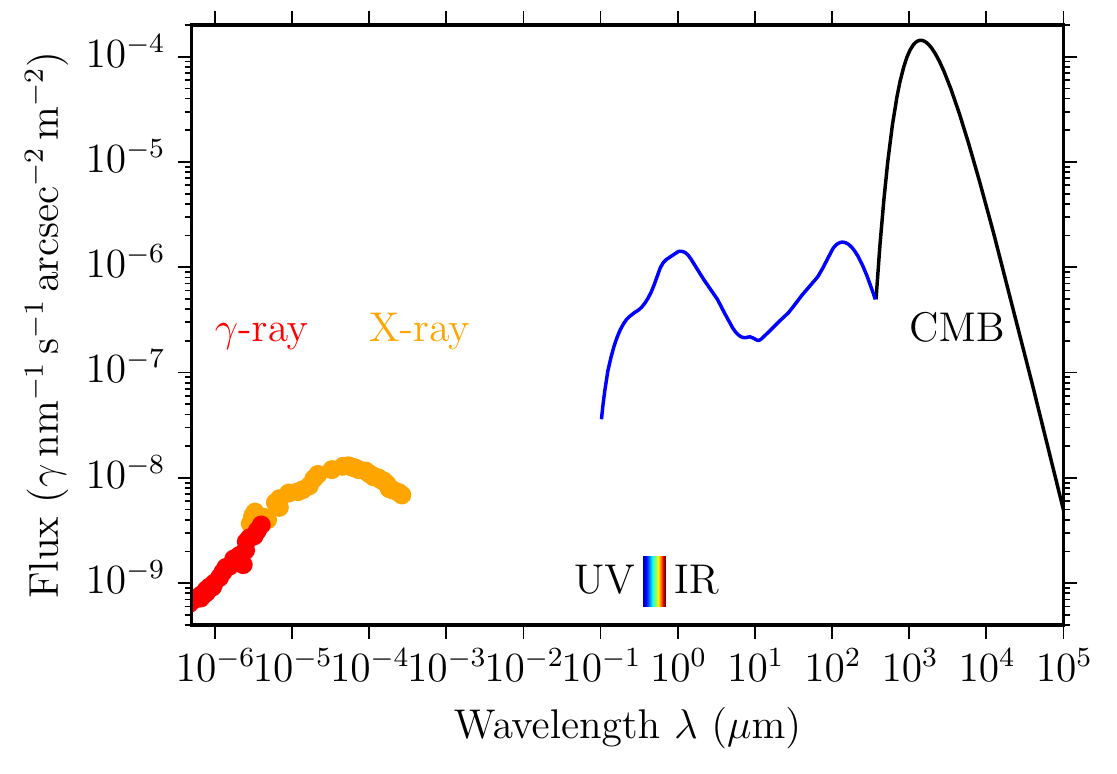}
\caption{\label{figure_background_sources}Figure from \citet{2017arXiv170603795H} on the intensity of the sky background after removal of the zodiacal light foreground. Data from \citet{1998ASPC..139...17L,2016RSOS....350555C,2016ApJ...827....6S}.}
\end{figure}

\section{Astrophysical noise}
\label{noise}
Noise photons impact the data rate through a reduction of the Holevo capacity, precisely in $N_{\rm M}$, which is the number of noise photons per mode. Noise can be instrumental and astrophysical, and we only discuss the latter here. Astrophysical noise strongly depends on wavelength, with peaks in the optical from nuclear fusion and in the FIR from re-radiated dust (Figure~\ref{figure_background_sources}). Due to absorption near the Lyman continuum (91.2\,nm), the UV/soft X-ray background is unknown. Noise levels for $\gamma$- and X-rays between keV and MeV are, within an order of magnitude, $10^{-8}\,\gamma$\,nm$^{-1}$\,s\,arcsec$^{-2}$\,m$^{-2}$, which is very low. Current technology would result in instrumental noise orders of magnitude higher. X-ray noise levels are two orders of magnitude below optical noise levels. Due to the logarithmic nature of capacity for signal-to-noise levels above unity, noise only impacts the data rate by a factor of a few.

\section{Discussion}
\label{discussion}

\begin{figure*}
\includegraphics[width=.5\linewidth]{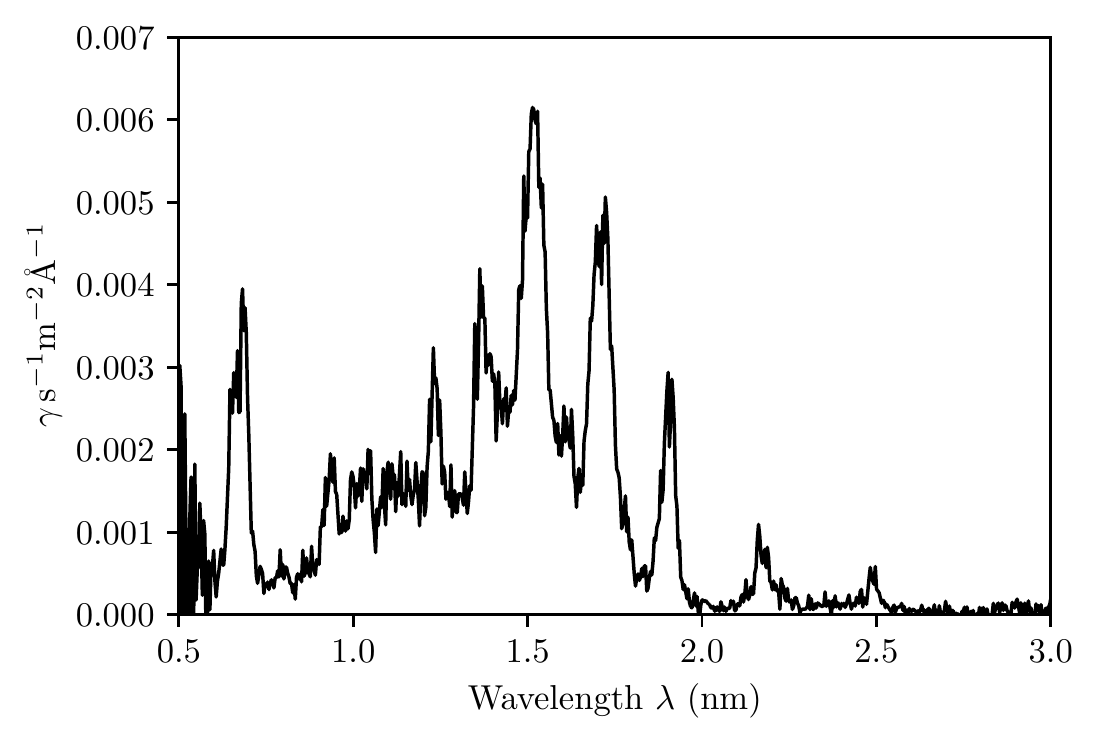}
\includegraphics[width=.5\linewidth]{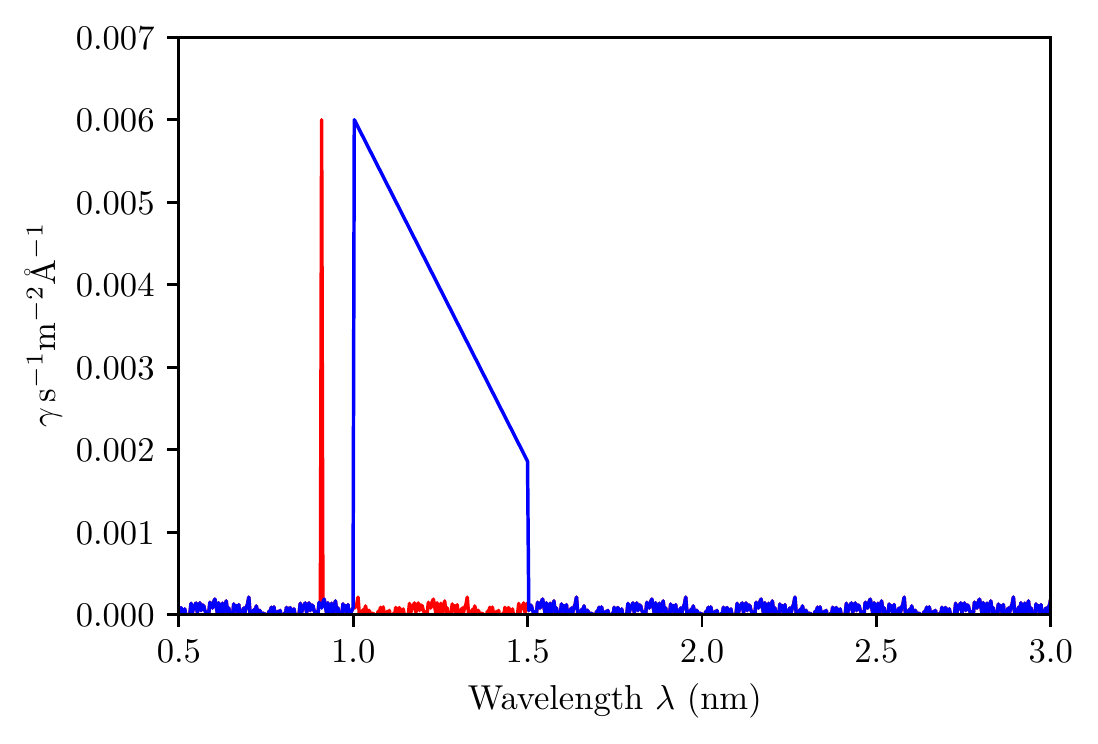}
\caption{\label{fig_xmm}Left: X-ray spectrum taken with XMM-Newton for Kepler's 1604 supernova remnant \citep{2004A&A...414..545C}. Right: Hypothetical spectra of monochromatic (red) and narrow-band (blue) X-ray signals.}
\end{figure*}

\subsection{Spectral specification of X-ray communication for future SETI searches}
The optimal spectrum for maximum data rate connection will have a hard cut at $\lambda>0.5$\,nm due to mirror surface roughness. The bandwidth (towards longer wavelengths) depends on the trade-off between the number of modes (more wavelength channels can encode more bits per photon, favoring large bandwidth) and the beam angle width (which increases due to diffraction with longer wavelength, favoring small bandwidth). As bitrate has a logarithmic relation to the number of modes in realistic cases, and the number of photons scales linear with beamwidth, the bandwidth will be small ($<100$\%). As a practical example, nanosecond time slots give $10^9$ modes per second, so that the monochromatic number of photons can be up to $10^8{\rm \,s}^{-1}$ before the mode penalty exceeds 1\% in bit rate. This means that a GB/s connection will be monochromatic if nanosecond technology is available. Such a connection over pc distances can be done with $D_{\rm t}=D_{\rm r}=10\,$m at MW power (section~\ref{data_rate}). 

This work suggests that if we wish to detect transmitters using optimal wavelengths for communication, we should consider carrying out SETI searches in the X-Ray regime, an idea that has been viewed in the past as quite exotic. The first idea to signal with X-ray pulses produced by high-altitude nuclear explosions was considered by \citet{1973cwei.book..398E}. The idea was later extended to the modulation of X-ray binaries \citep{1997JBIS...50..253C,2016arXiv160900330C} and dropping matter on neutron stars to produce X-ray flashes \citep{1977JBIS...30..112F,2015NewA...34..245C}.

X-ray communication is being considered for future deep space missions \citep{2014SPIE.9207E..16S}, inter-satellite communication \citep{2017SPIE10256E..1WM} and transmission during spacecraft reentry into earth's atmosphere \citep{2017JAP...121l3101L}.

We recommend expanding our SETI efforts towards targeted (at us) monochromatic (or narrow band) X-ray emission at 0.5-2 keV energies. Several current X-ray satellites cover this band: Swift (0.2--10\,keV), Chandra (0.1--10\,keV) and XMM-Newton (0.1--12\,keV), while INTEGRAL is only sensitive to higher energies (3\,keV--10\,MeV). Because of their designs with grazing incidence mirrors, all telescopes have small collecting areas (0.01\,m$^2$, 0.04\,m$^2$ and 0.45\,m$^2$ for Swift, Chandra and XMM, respectively). With a spectral resolution of $R=800$ beween 0.35--2.5\,keV, XMM is well equipped for the proposed narrow-band X-ray signals. As an example, we show the spectrum of Kepler's supernova remnant (SN1604) as observed with XMM-Newton by \citet{2004A&A...414..545C} in Figure~\ref{fig_xmm} (left panel). For comparison, synthetic monochromatic and narrow-band spectra are shown in the right panel. As there are thousands of archival X-ray spectra, a dedicated search for such features in existing data can be undertaken (Hippke et al. 2017 in prep.). Compared to the $5.7\times10^7$ sources visible to XMM over the entire sky \citep{2007arXiv0704.2293C}, the number of spectra is small ($<10^{-4}$ of all detectable sources), but an analysis will already place an upper limit on the number of bright optimally communicating (towards us) X-ray sources; at the moment this number is still unbound.

\subsection{Is perfect communication indistinguishable from blackbody radiation?}
It has been claimed that any sufficiently advanced technology is entirely indistinguishable from blackbody radiation \citep{2004AmJPh..72.1290L,2004PhRvA..69e2310G}. This is true for the case where an observer is not familiar with the encoding scheme used, and there is zero extinction between transmitter and receiver. For communication near $\lambda\approx$\,nm, the blackbody temperature would be $\approx3\times10^6$\,K, hotter than any stellar surface. Such communication would not be optimal for an isotropic beacon, however, because the energy per photon is unfavorably high compared to radio frequencies. 

First of all, it is a misconception that such a perfect blackbody radiation would be indistinguishable from natural sources. Almost all known luminous astrophysical bodies are \textit{not} perfect blackbodies, with the only exception being the \citet{1975CMaPh..43..199H} radiation of a BH\footnote{The peak wavelength of a BH is close to $16\times$ its Schwarzschild radius, resulting in a very low power of order $10^{-28}$\,W for a stellar mass BH.}. Even a star originally made of pure hydrogen produces heavier elements through nucleosynthesis, resulting in spectral absorption lines. A recent study used SDSS spectra and found 17 stars out of $789,593$ which resemble a perfect blackbody within their measuring uncertainty of 0.01\,mag$\approx$1\% \citep{2017arXiv171101122S}. The authors speculate that these objects are young DB white dwarfs with temperatures around $10000$\,K. High-resolution high-SNR spectroscopy should reveal some absorption lines; if this is not the case, these objects should be considered prime candidates for targeted OSETI. The detection of a perfect blackbody must be extraordinary evidence of artificial origin, as long as it is not a BH or a very young star without nucleosynthesis.

In practical communication, spreading the photons over the total spectrum like a blackbody is inefficient because the interstellar medium is less transparent at some frequencies, and these will be spared in an energy-efficient real-world communication. For example, even the nearest stars are dimmed by many magnitudes near the Lyman limit of 91.2\,nm, and one would avoid sending photons in and near this band, because these are (almost entirely) lost. Artificial communication would use hard cuts near such bands, in contrast to a continuous fade for natural sources. A spectrum would show rectangular absorption features instead of a smooth fade.

Also, the choice for a spread spectrum versus narrow band communication has marginal benefits for cases where the number of encoding modes is not the bottleneck. As an order-of-magnitude argument, narrow-band laser communication can readily employ $10^{10}$ modes by combining narrowband spectral plus timing encoding. Due to the logarithmic benefits of the number of modes, the number of photons received needs to exceed this order to make a spread spectrum useful -- otherwise the capacity (in bits per photon) is virtually identical, so that a blackbody spectrum yields virtually the same number of data bits (per photon and per unit time) as narrowband communication, and only complexity is added. 

Receiving $10^{10}$ photons per unit time (e.g., per second) is a considerable number, comparable to the average (isotropic) luminosity of the nearest stars ($10^{26}$\,W) per second per square meter as seen on earth. To justify the blackbody spread, this number needs to be exceeded by several orders of magnitude. Therefore, such a source would be prominent in the sky.

\subsection{Issues with current OSETI searches}
If interstellar communication is performed at nm wavelengths (keV energies), traditional optical SETI ($300 \dots 1000$\,nm) will not find it, even if it is directed at Earth. Also, Earth's atmosphere is intransparent for photons with $\lambda<291$\,nm, requiring space-based receivers.

The next issue is time resolution. Current OSETI experiments use ns ($10^{-9}$\,s) time resolution \citep{2000ASPC..213..545H,2009AsBio...9..345H,2017AJ....153..251T}. This is not a conscious choice, it is simply a technological limitation. Current (Earth 2017) technology offers superior timing on the transmitter side, compared to the receiver side, by at least 8 orders of magnitude.

Current photon-counting detectors can be relatively fast (timings below $10^{-10}$\,s) and efficient ($>90$\%) with a low dark count rate ($<1$\,c.p.s.), but suffer from long ($10^{-7}$\,s) reset times \citep{2013NaPho...7..210M}. Classical photomultiplier tubes offer timings (and reset times) of $10^{-9}$\,s \citep{2006NIMPA.563..368D}.

The shortest possible laser pulse length has decreased by 11 orders of magnitude during the last 50 years, from $100\,\mu$s in the free-running laser of \citet{1960Natur.187..493M} to 67 attoseconds \citep[$10^{-18}$\,s,][]{2012OptL...37.3891Z}. For a detailed history of the exponential improvements, see \citet{2004RPPh...67..813A}.

Capacity (in bits per photon) increases with short time slots, and the finite number of astrophysical noise photons per time slot decreases. Therefore, shorter pulses are preferred. If we have the technology to make such pulses, we can assume the same for ETI. So the obvious question is: What will a current OSETI ns receiver see, if the pulses are fs? The answer strongly depends on the flux levels.

To get a feeling for the numbers involved, consider the Harvard SETI's sensitivity threshold of 100 optical photons (with $\lambda=450 \dots 650$\,nm) per square meter, arriving at the telescope in a group within 5\,ns \citep{2004ApJ...613.1270H}. This translates into the fact that an (average) flux of $<2\times10^{10}$ photons per second per square meter can remain undetected, as long as the clustering appears in unresolved time slots. Even at a meager capacity of 1 bit per photon, the data rate could exceed Gbits/s and remain unnoticed because the encoding remains unresolved. For comparison, the flux from $\alpha$\,Cen~A is at the same level, $5\times10^{10}\,\gamma$\,sec$^{-1}$\,m$^{-2}$ \citep[distance 1.3 pc,][$L= 1.522 L_{\odot}$]{2016A&A...594A.107K}. Consequently, an isotropic emitter at pc distance with a flux level of $10^{27}$\,W and fs time slot communication would remain unnoticed, even if the OSETI telescope is aimed directly at it. As we can produce fs pulses today, it appears likely that ET communication timings are at least as short.

\section{Conclusion and outlook}
\label{conclusion}
Optimal directed communication (which maximizes data rates) within the laws of physics converges on $\lambda\approx$\,nm wavelengths (keV energies) due to the atomic surface smoothness limit, which cannot efficiently be surpassed with electromagnetic fields. Communication links using this technology can deliver high (Mbits/s, GBits/s) data rates over interstellar distances at moderate apertures and power. We may argue that mature civilizations would not use inefficient technologies such as burning kerosene or sending microwaves, optical or IR photons. Instead, we may assume that ET have developed cost-efficient technology required for the optimum communication, and have established such communication links. For these tight communication beams, the chance of a random intercept at Earth is very small \citep{2014JBIS...67..232F}, so that classical SETI is in vain. This topic will be covered in a subsequent part of this series.

\acknowledgments
\texttt{Acknowledgments. }
The authors are thankful to Silvia Vernetto for explanations on $\gamma$-ray absorption, 
to Cris More for an exchange on the similarity of information-efficient communication to black-body radiation, 
to Zachary Manchester and Ren\'{e} Heller for helpful discussions,
and to the Breakthrough Initiatives for an invitation to the Breakthrough Discuss 2017 conference at Stanford University.

\bibliographystyle{yahapj}

\end{document}